\documentclass{aastex}
\usepackage{emulateapj5,apjfonts,epsfig}

\shortauthors{KEMPNER \& SARAZIN}
\shorttitle{RADIO HALOS \& RELICS FROM WENSS}

\slugcomment{Accepted for publication in the Astrophysical Journal.}

\begin{document}

\title{Radio Halo and Relic Candidates from the Westerbork Northern Sky Survey}

\author{Joshua C. Kempner and Craig L. Sarazin}

\affil{Department of Astronomy, University of Virginia,
P. O. Box 3818, Charlottesville, VA 22903-0818}
\email{jck7k@virginia.edu,
cls7i@virginia.edu}

\begin{abstract}
We have undertaken a systematic search for diffuse radio halos and
relics in all of the Abell clusters that are visible in the Westerbork
Northern Sky Survey (WENSS).
In this survey we found 18 candidates, 11 of which are already known
from the literature, and 7 for which we provide the first evidence of
diffuse radio emission.
All the clusters in this sample show other evidence for a recent or
ongoing merger.
We also investigate the correlation between cluster X-ray luminosity
and radio power of halos.
We develop a very simple model for merger shocks that reproduces the
sense of this correlation, although it is probably not as steep as
the correlation in the data.
We discuss the implications of X-ray--radio correlations for future
detections of radio halos.
\end{abstract}


\section{Introduction} \label{sec:intro}

A number of clusters of galaxies are known to contain large-scale
diffuse radio sources which have no obvious connection to the clusters'
population of galaxies.
These sources are referred to as radio halos when they appear projected
on the center of the cluster, and are called relics when they are found
on the cluster periphery.
Because of projection effects, the distinction between halos and relics
is physically debatable, but does provide a convenient classification
for their discussion.
It is also possible that halos and relics in fact have different
physical origins, as we discuss below.
Halos (and relics; we will use ``halos'' hereafter for brevity) are
believed to be produced by synchrotron emission from a population of
relativistic electrons which have been accelerated or reaccelerated,
possibly by shocks in the intracluster medium
(e.g., Jaffe 1977; Roland 1981; Schlickeiser et al.\ 1987).
These shocks may be the product of cluster mergers.
In fact, all known halos are found in clusters which show other
signs of being in some stage of a merger
(e.g., Feretti 1999; Schuecker \& B\"{o}hringer 1999).
In the early stages of mergers, halos are often found on the border
between the subclusters, where the cluster gas is first being shocked
(e.g., Abell 85; Slee \& Reynolds 1984).
In more advanced mergers, more conventional centrally located halos
(e.g., Coma) and peripheral relics (e.g., Abell 3667) are found.

The most common explanation of the physical origin of halos and relics
is that they originate from particle acceleration in merger shocks, but
other theories have been suggested.
En{\ss}lin et al.\ (1998)
suggest that relics may trace shocks created in the initial structure
formation of the Universe.
Liang et al.\ (2000)
posits that turbulent reacceleration may maintain the population of
cosmic rays necessary to produce a halo after they have been accelerated
initially by merger shocks.
There is also disagreement over whether the cosmic ray electrons are
accelerated directly from the thermal population or are re-accelerated
cosmic rays previously produced in starbursts and AGN.
Alternately, the secondary electron model
(Dennison 1980; Blasi \& Colafrancesco 1999 and references therein)
proposes that the necessary electrons are created as a result of
interactions between cosmic ray protons and the intracluster gas.
The secondary electron model is unique in that it does not require
a merger to create a radio halo.

Halos and relics are rare phenomena, with only about 10--20 having been
known until quite recently.
With the completion of the
NRAO VLA Sky Survey
(NVSS: Condon et al.\ 1998)
and
Westerbork Northern Sky Survey
(WENSS: Rengelink et al.\ 1997),
systematic searches for additional sources have been possible, yielding
about 25 new halos and relics.
Evidence on the exact origin of halos and relics is still unclear,
mostly because of their small numbers.
With a larger sample, however we can begin to determine if the merger
shock picture of their formation is correct, and perhaps even
disentangle the different mechanisms for halo and relic formation, if
they are in fact different phenomena.

We assume $H_0 = 50$ km s$^{-1}$ Mpc$^{-1}$ and $q_0 = 0.5$ throughout
this paper.

\section{Sample and Source Selection} \label{sec:sample}

We conducted our search for new halos and relics using the publicly
available images of the WENSS.
The survey has an angular resolution of $54\arcsec \times 54\arcsec \csc
\delta$ at declination $\delta$, a typical noise level of 3.6~mJy beam$^{-1}$,
and covers the sky north of $\delta = 30^{\circ}$.

Our search was strictly limited to galaxy clusters in the ACO catalog
(Abell 1958; Abell, Corwin \& Olowin 1989) that fall within the region
covered by the WENSS.
This gave us a sample of 1001 clusters up to a redshift of $\sim 0.3$.
With the good $(u,v)$ coverage of the survey, the images were sensitive
to extended structures of up to $1^{\circ}$ in diameter.
We were therefore able to detect cluster-wide halos with a typical size of
1~Mpc at redshifts $z \gtrsim 0.01$, which covers the entire Abell
catalog.
Sources 1~Mpc in size do not become unresolved by the WENSS beam until a
redshift much greater than the limit of the Abell catalog.

We searched for diffuse sources using images from the WENSS on their own
as well as using these images overlaid on images from the
Digital Sky Survey (DSS).
Our criteria for a positive detection were that the sources had to {\it
a)}\/ have regions of surface brightness greater than the 2$\sigma$
level, {\it b)}\/ be resolved, {\it c)}\/ not be associated with an
optically identified galaxy, {\it d)}\/ not be clearly associated with a
known extended radio galaxy, and {\it e)}\/ not be an obvious blend of
unresolved sources.
We used images from the NVSS, and from the FIRST survey where available
(Becker, White, \& Helfand 1995)
to check for the existence of point sources.
In clusters within about $1^{\circ}$ of a very bright (peak-flux-to-rms
$\gtrsim 800$) source, uncleaned sidelobe structure had the potential
for causing confusion, but the large sizes of the WENSS images enabled
us to identify these cases and avoid confusion.
We further compared the WENSS images to images from the {\it ROSAT}
All-Sky Survey (RASS; Voges 1992; Tr\"umper 1993) for information about
the X-ray morphology of the clusters.

For sources which are resolved but smaller than about $1^{\circ}$,
the WENSS is surface brightness limited.
Figure~\ref{fig:sensitivity_sb} shows the mean surface brightnesses
of the sources in our sample as a function of redshift.
The dashed line is the nominal 1$\sigma$ surface brightness limit of WENSS
(3.6~mJy beam$^{-1}$).
The mean surface brightness was determined for each source by spreading
its integrated flux evenly over a convex region which just encloses the
2$\sigma$ surface brightness contours for that source.
In reality, however, few of the sources fill this region completely, and
most of them have splotchy surface brightness distributions with peaks
which are much higher than this average value.
This is the reason that our survey limit on the mean surface brightness
appears to be be slightly below the WENSS noise limit.
Note that our cluster survey would not be sensitive to sources which
are very close or very large ($\ga$$1^{\circ}$) or sources which are
very small or very distant, which would be unresolved ($\la$1\arcmin).

\centerline{\null}
\vskip2.55truein
\includegraphics{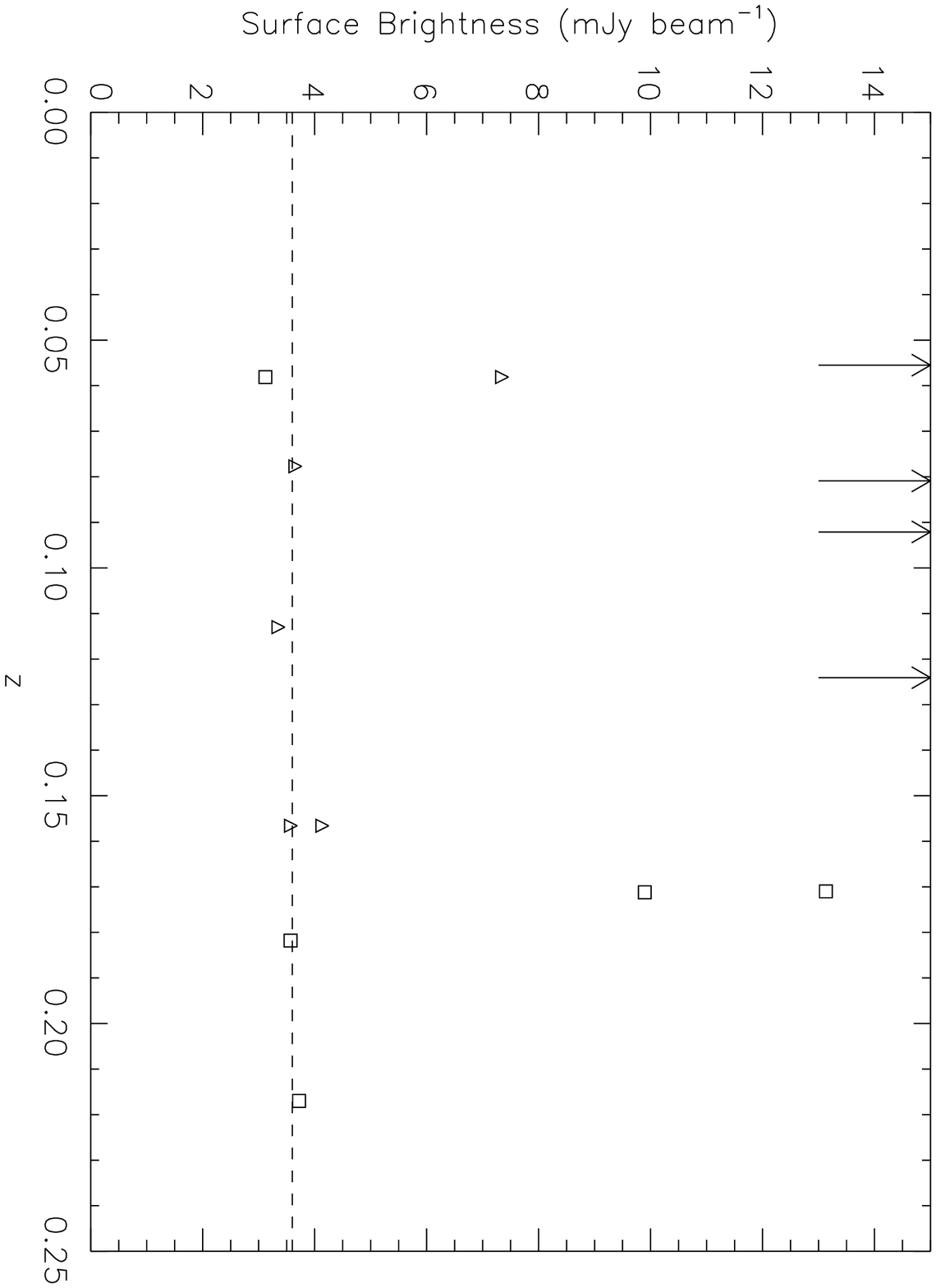}
\figcaption{Mean surfaces brightnesses and redshifts are shown
for the detected sources, with halos indicated by squares and relics by
triangles.
The mean surface brightness assumes that the flux is distributed over a
convex region which just encloses the 2$\sigma$ surface brightness
contours.
The nominal 1$\sigma$ surface brightness limit of the WENSS 
(3.6 mJy beam$^{-1}$) is indicated by the dashed line.
\label{fig:sensitivity_sb}}

\vskip0.2truein

\section{Halo and Cluster Candidates} \label{sec:candidates}

Using the criteria detailed above, we found a sample of 16 clusters
containing candidate halos or relics.
Of this sample, 7 were not known prior to this study, while 9 had
previously been discovered.
Of the previously known halos and relics which fell into our sample, all
but one (Abell~2218, see \S\ref{sec:notes}) were detected in this study.
Abell~2218 is included in our analysis for completeness.
The halo in Abell~2218 was not detected because it was nearly unresolved 
in WENSS.
Five of the seven new detections are associated with clusters which are
not part of the Ebeling et al.\ (1996) catalog of X-ray bright clusters,
and therefore were not in the NVSS sample (Giovannini et al.\ 1999) that
comprises about half of the previously known halos and relics in our
sample.

Table~\ref{tab:candidates} lists the candidate clusters.
For each cluster, we list the name, redshift $z$, position, X-ray
luminosity in the 0.1--2.4 keV rest frame energy range,
Rood-Sastry (1971) class, Bautz-Morgan (1970) class, radio morphology,
and whether the cluster halo/relic had been detected previously.
Data in columns 2, 3, and 4 are taken from Ebeling et al.\ (1996)
except where noted.
The X-ray luminosities in column 5 came from the literature
(sources noted), except for A~796, where the luminosity was determined
from the RASS flux.
Column 8 gives the morphology of the radio source: ``H'' for a halo,
``R'' for a relic, and ``u'' for uncertain.
According to convention, we classified diffuse sources as halos if they
were centered on the cluster and as relics if they appeared on the
cluster periphery.
Sources whose status as a halo or relic are uncertain were labeled as
such.
These include faint sources, sources which may be a blend of unresolved
sources, and sources which have tentative but not obvious associations
with optical sources.
Uncertain sources comprise about half of our sample of new detections.

\begin{table*}[hbt]
\caption{Properties of Clusters Containing Halo and/or Relic
Candidates \label{tab:candidates}}
\begin{center}
\begin{tabular}{lccccllcc}
\hline
\hline
Name &
$z$ &
RA (J2000) &
Dec. &
$L_X$ (0.1--2.4 keV) &
RS\tablenotemark{a} &
BM\tablenotemark{b} &
Radio &
Previously \\
 &
 &
(h~\phn m~\phn s)~\phn &
$\phn(^{\circ}~\phn \arcmin~\phn \arcsec)$ &
($10^{44}$ erg s$^{-1}$) &
 &
 &
Morphology\tablenotemark{c} &
Detected? \\
\hline
A 665   & 0.1818     & 08 30 47.4 & +65 51 14     & 14.78$^6$ & I c & III:    & H   & y \\
A 697   & 0.282$^2$  & 08 42 57.6 & +36 21 59     & 16.30$^5$ & F   & II-III  & u   & n \\
A 725   & \phn0.0921$^7$ & 09 01 10.1 & +62 37 20     & \phn0.80$^1$  & I c & \nodata & R   & n \\
A 773   & 0.2170     & 09 17 54.0 & +51 42 58     & 12.35$^6$ & B   & II-III: & H   & y \\
A 786   & \phn0.1241$^8$ & 09 28 49.7 & +74 47 55     & \phn1.53$^1$  & F   & \nodata & R   & y \\
A 796   & 0.1475\tablenotemark{e} & 09 28 00.0 & +60 23 00     & 1.38      & C   & III     & u   & n \\
A 1240  & 0.1566\tablenotemark{e} & 11 23 37.6 & +43 05 52$^3$ & \phn1.36$^3$  & C   & III     & R+R & n \\
A 1452  & \phn0.0630$^8$ & 12 03 38.8 & +51 44 18     & \nodata   & C   & \nodata & u   & n \\
A 1758a & 0.2800     & 13 32 45.3 & +50 32 53     & 11.22$^4$ & F   & \nodata & u   & y \\
A 1914  & 0.1712     & 14 26 02.2 & +37 50 06     & 17.93$^6$ & L   & II:     & H   & y \\
A 2034  & 0.1130     & 15 10 11.7 & +33 29 12     & \phn6.86$^4$  & L   & II-III: & R   & n \\
A 2061  & 0.0777     & 15 21 17.0 & +30 38 24     & \phn3.92$^4$  & L   & III:    & R   & n \\
A 2218  & 0.1710     & 16 35 52.8 & +66 12 59     & \phn8.77$^6$  & C   & II:     & H   & y \\
A 2219  & 0.2281     & 16 40 22.5 & +46 42 22     & 19.80$^4$ & F   & III     & u   & y \\
A 2255  & 0.0809     & 17 12 45.1 & +64 03 43     & \phn5.68$^6$  & C   & II-III: & H   & y \\
A 2256  & 0.0581     & 17 04 02.4 & +78 37 55     & \phn6.99$^6$  & B   & II-III: & H+R & y \\
A 2319  & 0.0555     & 19 21 05.8 & +43 57 50     & 12.99$^6$ & cD  & II-III  & H   & y \\
\hline
\end{tabular}
\end{center}
\tablenotetext{a}{RS types are as follows: cD = single cD galaxy;
B = ``binary,'' two dominant galaxies; F = ``flat,'' no dominant galaxy;
C = ``clumpy'' spatial distribution of galaxies; L = ``linear''
distribution of galaxies; I = ``irregular'' galaxy distribution.}
\tablenotetext{b}{Bautz-Morgan types are as follows: type II clusters have
no cD galaxy but have one or more Virgo-type giant ellipticals; type
III clusters have no dominant galaxies; type II-III are intermediate;
a colon after the type indicates an uncertain type estimate.}
\tablenotetext{c}{Radio morphology notation is: H = halo; R = relic;
u = uncertain.}
\tablenotetext{e}{Redshift estimated using the method described by
Ebeling et al.\ (1996); these clusters' positions are taken from Abell
(1958) except where another reference is indicated.}
\tablerefs{(1) B\"{o}hringer et al.\ 2000;
(2) Crawford et al.\ 1995;
(3) David, Forman, \& Jones 1999;
(4) Ebeling et al.\ 1996;
(5) Ebeling et al.\ 1998;
(6) Feretti 2000;
(7) Owen, Ledlow, \& Keel 1995;
(8) Struble \& Rood 1987}
\end{table*}

In Table~\ref{tab:radio}, we list the properties of the halo and relic
candidates themselves.
Clusters with multiple sources have one entry for each source.
Columns 2 and 3 list the WENSS and NVSS fluxes, respectively, as
determined by the WENSS 2$\sigma$ surface brightness contours.
In clusters where point sources were superposed on the likely extent of
the diffuse emission, we could not satisfactorily subtract off those
sources due to substructure in the beam.
Instead, we masked out the sources and added a systematic error to our
measurement which assumes that the surface brightness in the masked
regions is equal to the average in the unmasked area.
This may in some cases underestimate the actual error if sidelobes from
these sources are present in the unmasked area, or if the surface
brightness in the masked region differs significantly from the mean.
Col.\ 4 gives the spectral index $\alpha$ derived from these fluxes.
Here, the radio power $P_\nu$ varies with frequency as
$P_\nu \propto \nu^\alpha$.
Col.\ 5 lists the monochromatic radio power of the source at 327~MHz in
the cluster rest frame.
The derived spectral index is used to correct for the redshift, assuming
that the same spectral index holds for lower observed frequencies.
The position in columns 6 and 7 is the estimated center of the diffuse
source.
Columns 8 and 9 list the largest angular size $LAS$ of the
source and the corresponding largest linear size $LLS$ at the redshift of
the cluster, both evaluated at 327 MHz.
Column 10 gives the projected distance of the source from the cluster
center (see Table~\ref{tab:candidates}) for relic sources only.

\begin{table*}
\caption{Properties of Halo and Relic Candidates\label{tab:radio}}
\begin{center}
\begin{tabular}{@{}lccccccccc@{}}
\hline
\hline
Name &
$S_{327}$ &
$S_{1400}$ &
$\alpha$ &
$P_{\nu}$ &
RA (J2000) &
Dec. &
$LAS$ &
$LLS$ &
Distance \\
&
(mJy) &
(mJy) &
&
($10^{24}~{\rm W~Hz^{-1}}$) &
(h~\phn m~\phn s) &
($^{\circ}~\phn\arcmin$) &
(\arcmin) &
(kpc) &
(Mpc) \\
\hline
A 665   & $108 \pm 17$   & $16 \pm 2$ & $-1.30 \pm 0.25$ &    $17.61 \pm 0.10$ &
08 30 58 & +65 50 & \phn8.0 & 1900 & \nodata \\
A 697   & $29 \pm 6$     & $\phn7 \pm 2$  & $-0.97 \pm 0.28$ &    $11.16 \pm 0.18$ &
08 42 56 & +36 22 & \phn2.9 & \phn920 & \nodata \\
A 725   & $76 \pm 9$     & $\phn6 \pm 1$  & $-1.73 \pm 0.35$ & $\phn3.10 \pm 0.18$ &
09 01 29 & +62 38 & \phn3.2 & \phn440 & 0.31 \\
A 773   & $35 \pm 7$     & $\phn8 \pm 1$  & $-1.02 \pm 0.26$ & $\phn7.85 \pm 0.11$ &
09 18 04 & +51 42 & \phn5.3 & 1400 & \nodata \\
A 786   & $319 \pm 22$   & $104\pm 3\phn$ & $-0.77 \pm 0.06$ &    $21.85 \pm 0.04$ &
09 22 16 & +75 00 & \phn8.2 & 1400 & 5.0\phn \\
A 796   & $\phn53 \pm 15$    & $\phn8 \pm 3$  & $-1.34 \pm 0.60$ & $\phn5.55 \pm 0.30$ &
09 27 38 & +60 23 & \phn6.1 & 1200 & \nodata \\
A 1240N & $32 \pm 7$     & $\phn8 \pm 1$  & $-0.96 \pm 0.26$ & $\phn3.61 \pm 0.12$ &
11 23 28 & +43 10 & \phn4.0 & \phn850 & 1.3\phn \\
A 1240S & $\phn54 \pm 10$    & $11 \pm 2$ & $-1.11 \pm 0.27$ & $\phn6.22 \pm 0.13$ &
11 23 47 & +43 01 & \phn4.6 & \phn960 & 1.5\phn \\
A 1452  & $\phn54 \pm 15$    & $14 \pm 3$ & $-0.92 \pm 0.32$ & $\phn0.95 \pm 0.17$ &
12 03 18 & +51 45 & \phn7.9 & \phn780 & \nodata \\
A 1758  & $\phn55 \pm 11$    & $11 \pm 2$ & $-1.13 \pm 0.31$ &    $21.52 \pm 0.14$ &
13 32 44 & +50 32 & \phn4.3 & 1300 & \nodata \\
A 1914\tablenotemark{a}&$114 \pm 29$&$20 \pm 3$&$-1.19 \pm 0.34$&$16.07 \pm 0.13$ &
14 25 58 & +37 48 & \phn6.8 & 1500 & \nodata \\
A 2034  & $44 \pm 9$     & $\phn8 \pm 2$  & $-1.17 \pm 0.36$ & $\phn2.60 \pm 0.19$ &
15 10 17 & +33 31 & \phn5.7 & \phn920 & 0.35 \\
A 2061  & $104 \pm 15$   & $19 \pm 3$ & $-1.17 \pm 0.23$ & $\phn2.85 \pm 0.12$ &
15 20 06 & +30 29 & \phn7.7 & \phn920 & 2.1\phn \\
A 2218  & $\phn9 \pm 4$      & $\phn\phn\phd1 \pm 0.6$& $-1.46 \pm 1.04$ & $\phn1.33 \pm 0.53$ &
16 35 46 & +66 12 & \phn1.5 & \phn340 & \nodata \\
A 2219  & $19 \pm 6$     & $\phn2 \pm 1$  & $-1.44 \pm 0.82$ & $\phn5.27 \pm 0.44$ &
16 40 11 & +46 44 & \phn2.9 & \phn810 & \nodata \\
A 2255\tablenotemark{a}&$360 \pm 44$&$18 \pm 5$&$-2.06 \pm 0.57$&$11.47 \pm 0.37$ &
17 12 54 & +64 04 & \phn7.6 & \phn930 & \nodata \\
A 2256R\tablenotemark{a}&$1165 \pm 107$&$190 \pm 19$&$-1.25 \pm 0.17$&$17.69 \pm 0.08$ &
17 02 57 & +78 43 & 15.8 & 1450 & 0.66 \\
A 2256H\tablenotemark{a}&$126 \pm 32$&\nodata& \nodata   & $\phn1.99 \pm 0.60$\tablenotemark{b} &
17 04 46 & +78 39 & \phn7.6 & \phn700 & \nodata \\
A 2319\tablenotemark{a}&$204 \pm 40$&$32 \pm 6$&$-1.28 \pm 0.35$&$\phn2.83 \pm 0.16$&
19 21 11 & +43 56 & \phn6.6 & \phn580 & \nodata \\
\hline
\end{tabular}
\end{center}
\tablenotetext{a}{Discrete sources superposed on diffuse emission have
been masked out as described in the text.}
\tablenotetext{b}{Power calculated by assuming the spectral index
of $-1.9$ given by Costain et al.\ (1972) with an associated error of
$\pm 0.3$.}
\end{table*}

In the cases of many of the well studied and more diffuse sources, they
are more extended than we observe them to be in either the WENSS
or the NVSS.
These sources contain very low surface brightness emission around their edges,
and what we see are only the brighter cores.
Therefore both our size measurements and our flux measurements are known
to be underestimated in some cases and may be in others as well.

WENSS and NVSS radio contours of the clusters in Table~\ref{tab:candidates}
are shown in Figure~\ref{fig:images}, superposed on DSS optical images.
The images are of variable size, although most are approximately
12\arcmin$\times$12\arcmin.
All include the cluster center listed in Table~\ref{tab:candidates} and
extend to include the radio emission of interest.

\section{Notes on Individual Sources} \label{sec:notes}

{\it A 665.}\/
The halo is detected at better than the $6\sigma$ level in the WENSS,
only slightly less significantly than in the NVSS.
The spectral index we find is consistent with the lower limit of
$\alpha < -0.6$ found by Moffet \& Birkinshaw (1989).

{\it A 697.}\/
The diffuse emission is located in the cluster center and extends to the
NW.
The peak of the emission is roughly coincident with the X-ray centroid
of the cluster.
The absence of point sources in this region in the FIRST survey and the
lack of bright ellipticals in the DSS suggest that the diffuse emission
is in fact real, despite its low flux.
Because of the large uncertainties in its flux, however, we will consider
it to be uncertain until deeper, more detailed observations can be made.

{\it A 725.}\/
The brightest radio source in the cluster is associated with the bright
elliptical at the cluster center.
The relic is seen as an arc of diffuse emission to the NE of this source.
The X-ray gas as seen in the RASS is slightly elongated along the axis
connecting the relic and the cluster center.
It has one of the steepest spectra of any source in our sample.

{\it A 773.}\/
The diffuse emission is quite splotchy and irregular in the WENSS image
but is detected at the 5$\sigma$ level.
Its existence had been previously known from the NVSS
(Giovannini et al.\ 1999).

{\it A 786.}\/
This relic (upper right corner in Figure~\ref{fig:images}) is at an
unusually large projected distance ($\sim 5$ Mpc) from the center of the
cluster.
Its association with the cluster is based on the relic being coincident
in projection with two galaxies of similar redshift to that of the
cluster.
The cluster is a member of Rood Group \#27 (Rood 1976), although the
other Abell clusters in the group lie to the southwest of A 786.
Based on the presence of the galaxies at the position of the relic, the
supercluster also is likely to extend to the northwest.

The cluster's X-ray morphology indicates that it is a double cluster,
with the position given in Table~\ref{tab:candidates} lying between the
two X-ray clusters.

The spectral index we measure is consistent with the value determined by
Harris et al.\ (1993).

{\it A 796.}\/
The centrally located diffuse emission in this cluster is quite large
and has very low surface brightness.
Deeper imaging will be needed to confirm the presence of a halo in this
cluster.

{\it A 1240.}\/
Two roughly symmetric relics are found in this cluster. 
They are indicated in Table~\ref{tab:radio} by their positions relative
to the cluster center, north or south.
They appear to either side of the cluster center at projected distances
of $\sim 1.3$ and $\sim 1.5$~Mpc.
To compare the radio morphology with that of the X-ray emitting
intracluster gas, we extracted a pointed {\it ROSAT} PSPC observation
from the archive. 
This 11.9 ksec exposure (RP900383) was aimed at another target, and
A~1240 is located about 27\arcmin\ from the center of the field.
As a result of the poorer angular resolution and supporting rib
structure at this location in the detector, the image is of lower
quality than if the field had been centered on A~1240.
The X-ray image was corrected for particle background and exposure using
the HEASARC ftools software package.
The resulting X-ray image in the R4R7
(roughly, 0.5-2.0 keV) band
was adaptively smoothed with a minimum signal-to-noise ratio of
3 per smoothing beam.

\setcounter{figure}{2}
\centerline{\null}
\vskip3.0truein
\includegraphics{fig3.eps}
\figcaption{Adaptively smoothed {\it ROSAT} PSPC image of Abell
1240 (greyscale) with superposed WENSS image contours.  The greyscale is
a square root scaling and ranges from 0 to $2 \times 10^{-4}$ counts
sec$^{-1}$ pixel$^{-1}$; the pixels are 15\arcsec$\times$15\arcsec.  The
radio image is shown in 2- and 3.5-$\sigma$ surface brightness contours.
\label{fig:A1240}}
\vskip0.2truein

The smoothed image is shown in Figure~\ref{fig:A1240}, superposed
with the WENSS contours.
The {\it ROSAT} observation of the cluster show a double X-ray morphology
consistent with a slightly asymmetric merger.
Both of the radio relics are
located on the edge of the cluster X-ray gas;
both have similar luminosities and spectra.
The axis of symmetry of the relics is roughly aligned with the apparent
axis of the merger.
Hydrodynamic simulations of off-center mergers (Ricker \& Sarazin 2000)
show a similar slight misalignment of the axis connecting the merger
shocks with that of the cluster centers.
Although projection effects introduce some uncertainty, the location of
the relics is consistent with the expected location of merger shocks.
This cluster joins Abell 3667 and Abell 2345 in a growing class of
clusters with symmetric double radio relics.

{\it A 1452.}\/ 
This possible halo is quite large, but not very powerful.
It has relatively low surface brightness and is fairly splotchy.
Deeper imaging is needed to confirm the presence of diffuse emission in
this cluster.

{\it A 1758a.}\/
The brightest source is identified with a Narrow Angle Tail galaxy
(O'Dea \& Owen 1985) with the tail pointing to the SE.
The bulk of the remaining emission is resolved into two sources in the
FIRST image, but the faint emission to the south of these and between
these and the tailed galaxy may be diffuse emission unrelated to the
point sources, or it may be a blend of fainter point sources.
The presence of diffuse emission is therefore considered uncertain.

{\it A 1914.}\/ 
A very steep spectrum source has been known to exist in this cluster for
quite some time (e.g., Kulkarni et al.\ 1990).
The center of this cluster contains a number of point sources visible in
the FIRST images, at least two of which overlap the radio halo.
Accurately masking the point sources in the WENSS image was difficult
due to their small separation, so the actual error in our measurement
may be slightly greater than that quoted in Table~\ref{tab:radio}.

{\it A 2034.}\/
The extended emission is located north of the cluster center.
It is coincident with a discontinuity in the X-ray surface brightness
which may indicate the presence of a merger shock.
Forthcoming observations of the cluster with XMM-Newton and Chandra will
be able to confirm the existence of a merger shock at this location.

{\it A 2061.}\/
The diffuse emission in this cluster is visible as two closely separated
sources $\sim 18\arcmin$ SW of the cluster center.
The two regions are most likely part of a single radio relic.
The FIRST survey shows two point sources: one just south of the relic,
the other just north.
This second source is too faint to be clearly visible in the WENSS, but
is obvious in the NVSS.

As the cluster's RS type indicates, it is quite elongated, with the
position angle of the galaxy distribution extending in the direction of the
relic.
The X-ray gas also has a bimodal distribution
(Crawford et al.\ 1999).
The RASS image shows that this extension has the same orientation as
that of the optical galaxy distribution.

{\it A 2218.}\/
This fairly small halo was first detected by Moffet \& Birkinshaw
(1989).
It is nearly unresolved in WENSS and is quite faint, causing it to be
missed by our survey, but we include it here for the sake of
completeness.
This was the only halo or relic found in the literature which was part
of our initial sample but which we failed to detect in the WENSS.

{\it A 2219.}\/ 
This possible halo was first discovered in the NVSS
(Giovannini et al.\ 1999)
but was considered uncertain.
The WENSS data improve the situation somewhat, as the halo candidate
appears at slightly better than the 3$\sigma$ level as an arc of
emission to the west and NW of the cluster.
The emission to the SW that is visible in the NVSS appears to be a blend
of point sources which are resolved by the FIRST survey, but these
sources cannot explain the diffuse emission to the NW.
Since the halo candidate was still not detected with a high signal-to-noise,
we continue to consider this candidate as uncertain.

{\it A 2255.}\/ 
The spectral index we measure is consistent with the integrated
$\alpha \sim -1.7$ derived by Feretti et al.\ (1997a).
The somewhat steeper value we find is probably caused by including the
``hole'' region in which there is no significant emission at 20~cm.
This cluster also contains a relic (e.g., Burns et al.\ 1995) which is
not visible in either the WENSS or NVSS images at a level which is
considered significant.

{\it A 2256.}\/
This cluster contains perhaps the most spectacular diffuse radio source
in our sample. 
It contains a large radio relic to the west of the cluster.
We find a steeper spectral index for the relic than that reported by
R\"ottgering et al.\ (1994),
although the increased sampling of short $(u,v)$ spacings in the WENSS
as compared to their observations at 90~cm with the VLA B-array probably
accounts for the discrepancy.

We also observe a central radio halo which was detected by Bridle \&
Fomalont (1976) at 610~MHz, and which is easily visible in the WENSS.
Because of its very low surface brightness (it is rarely brighter than 2
mJy beam$^{-1}$ in the WENSS image) and its ultra-steep spectrum, this
halo has largely been ignored by subsequent studies of A 2256, which
have instead focused on the relic and the numerous head-tail galaxies in
the cluster.  Costain et al.\ (1972) found the spectral index of A 2256
between 22.25 and 81.5~MHz to be $-1.9$, suggesting the presence of a
centrally located ultra-steep spectrum source.  Their data also suggest
that this source is of comparable size to, or slightly bigger than the
observed extent of the halo at 327~MHz.  Based on this steep spectral
index and the halo's flux at 610~MHz (Bridle \& Fomalont 1976), we would
expect a flux more than 2.5 times what we observe at 327~MHz.  However,
since the halo is quite faint and is perhaps slightly larger than is
evident from the WENSS image, our measurement should be taken as a lower
limit to the actual flux of the halo.  The flux of the halo at 1400~MHz
is expected to be below the sensitivity of the NVSS, and indeed it is
not seen there at all.

The halo and relic measurements are listed separately in
Table~\ref{tab:radio} and are named accordingly.

{\it A 2319.}\/
This halo was studied extensively by Feretti et al.\
(1997b).
It is much more extended than it appears in either the WENSS or
NVSS, so much of the total flux in the halo is missing in our
measurements from both surveys.

\section{The Radio--X-ray Correlation} \label{sec:disc}

\subsection{Correlations for Clusters with Radio Detections}
\label{sec:disc_radio}

We now consider the correlation between the radio power of a cluster
halo or relic and the cluster's other properties.
We begin by limiting our statistical analyses to the radio measurements
done in this study, because of the difficulty of developing a consistent
sample of low-frequency radio data on radio halos and relics based on
previous observations.
Most previous large surveys of these objects have been done at 20~cm. 
We then discuss correlations at higher frequencies and attempt to apply
them to our 327~MHz data.

We find no significant correlation between the radio power of the source
and the RS type or the Abell Richness class of the cluster.
The distribution of halo/relic candidates in redshift space is uniform,
but given the small number of clusters in which we detect diffuse
radio emission, this result is probably not meaningful.

\centerline{\null}
\vskip2.45truein
\includegraphics{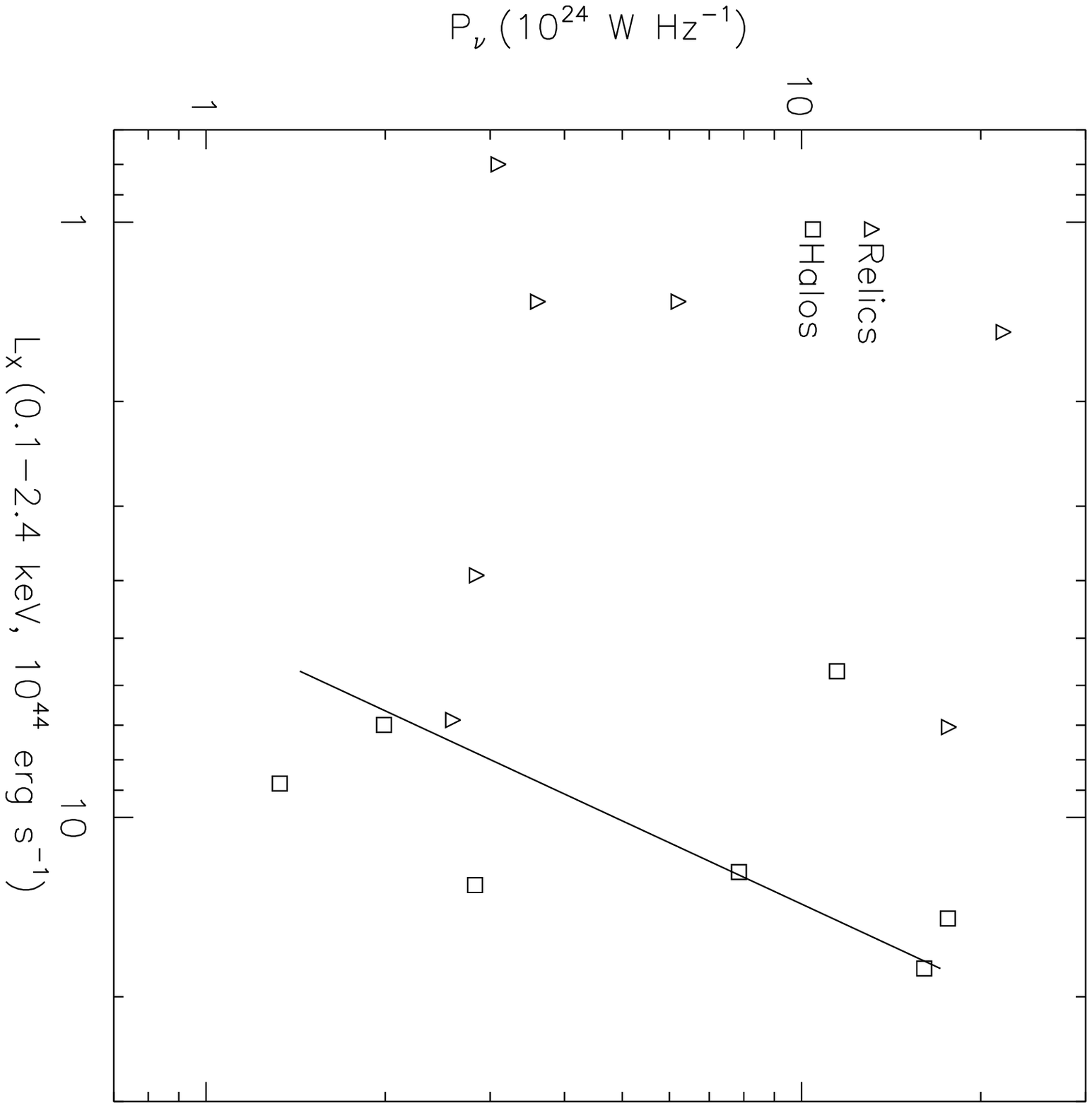}
\figcaption{Monochromatic radio power at 327~MHz in the cluster
rest frame plotted against X-ray luminosity in the 0.1--2.4 keV rest
frame band.
The solid line is the best fit to the halo data, using a
fixed logarithmic slope of 2.15 derived from the 20~cm data in
Feretti (2000).
\label{fig:P-L}}
\vskip0.2truein

We do not find a correlation between radio power and X-ray luminosity
when halos and relics are considered together
(Figure~\ref{fig:P-L}).
When considered separately, we find a correlation coefficient of $r =
0.56$ for halos, consistent with no correlation.
This may, however, be a result of missing flux from having few short
baselines, as more complete data at 1.4~GHz
(Feretti 2000; Liang et al.\ 2000)
do show a correlation.
Relics show no sign of a correlation either, although the X-ray data for
these clusters is quite poor.
It should be noted that the clusters in which we find the first evidence
of diffuse radio emission all display relatively faint halos or relics
and generally have X-ray luminosities which are comparable to or
slightly fainter than those of clusters already known to have halos or
relics.
With better radio data for these sources, they may extend the
correlation found at 1.4~GHz.

\subsection{A Simple Merger Model for the Radio--X-ray Correlation}
\label{sec:disc_model}

We now consider a very simple analytic model for the radio luminosity as a
function of X-ray luminosity, on the assumption that the radio emitting
electrons are accelerated in cluster mergers.
In general, a cluster of mass $M$ may undergo mergers with clusters or
groups having any mass $M^\prime \le M$.
(We take the mass $M$ to apply to the larger subcluster in the merger.)
However, mergers with much smaller subclusters are unlikely to produce
very strong shocks with large energy fluxes.
Thus, we consider only major mergers with $M^\prime \approx M$, and
for simplicity we assume that the mergers are symmetric ($M^\prime = M$).
Mergers occur with a distribution of impact parameters $b$, which
can affect the strength of the shocks produced
(e.g., Ricker \& Sarazin 2000).
To simplify the model, we assume that the mergers occur with $b = 0$
(zero impact parameter, or a head-on merger).
A distribution of values of $M^\prime$ and $b$ should introduce dispersion
into the correlation between X-ray and radio luminosities, but should
not affect the correlation itself.
Let us characterize each of the subclusters by the mass $M$, characteristic
radius $R$, and velocity dispersion $\sigma$.

To begin with, we assume that the radio emitting electrons have short
lifetimes and are accelerated by cluster merger shocks.
Radio observations of Galactic supernova remnants indicate that
shocks at velocities similar to those in cluster mergers convert
$\sim$3\% of the shock energy flux into accelerating relativistic
electrons
(e.g., Blandford \& Eichler 1987).
Thus, we will assume that the rate as which energy is deposited by
accelerating cosmic ray electrons $\dot{E}_{CR}$ is
\begin{equation} \label{eq:edot_cr}
\dot{E}_{CR} = f_{CR} \dot{E}_{shock} \, ,
\end{equation}
where $\dot{E}_{shock}$ is the rate at which energy is being deposited
in merger shocks, and
$f_{CR} \sim 0.03$ is the fraction of the shock energy that goes into
accelerating cosmic rays, and is assumed to be constant.
Assume the two subcluster approach with a velocity $v$.
Then, the rate of energy deposition by shocks is approximately
\begin{equation} \label{eq:e_shock}
\dot{E}_{shock} \approx \frac{1}{2} \rho_{gas} v^3 A \, ,
\end{equation}
where $\rho_{gas}$ is the pre-shock density of the thermal gas in the
intracluster medium (ICM),
and $A \sim \pi R^2 $ is the cross-sectional area of the merger shocks.
The merger velocity of subclusters is determined by their infall
velocity from the turnaround radius in the Hubble flow, and is close to
the escape velocity.
Thus, $v^2 \sim G M / R$.

The average density of the thermal gas is
\begin{equation} \label{eq:rho_gas}
\rho_{gas} = \frac{M_{gas}}{4 \pi R^3 / 3 } \, .
\end{equation}
where $M_{gas}$ is the total gas mass in the cluster.
We write the gas mass as
\begin{equation} \label{eq:m_gas}
M_{gas} = f_{gas} M \, ,
\end{equation}
where $f_{gas}$ is the cluster gas fraction.
The X-ray observations of rich clusters indicate that
$f_{gas} \approx 0.2 \approx$ constant
(e.g., Arnaud \& Evrard 1999).

We will assume that the magnetic field strength $B$ in the cluster radio
source is $\la$3 $\mu$G, so that the main loss process for the radio
emitting electrons is inverse Compton (IC) scattering of cosmic
microwave background photons.
Then the lifetime of an electron with an energy of
$\gamma m_e c^2$ is
(e.g., Sarazin 1999)
\begin{equation} \label{eq:t_IC}
t_{IC} = 7.7 \times 10^{9} \, \left( 1 + z \right)^{-4}
\left( \frac{\gamma}{300} \right)^{-1} \,
{\rm yr}
\, .
\end{equation}
The average frequency of synchrotron radiation
produced by an electron with a Lorentz factor of $\gamma$ is
$\langle \nu_{rad} \rangle = (55/96) (\sqrt{3}/\pi)
\gamma^2 (e B / m_e c) \sin \theta$, where $\theta$ is the pitch angle
of the electron
(e.g., Rybicki \& Lightman 1979).
This gives
\begin{equation} \label{eq:nu_rad}
\langle \nu_{rad} \rangle \la
560
\left( \frac{\gamma}{10^4} \right)^2
\left( \frac{B}{1 \, \mu{\rm G}} \right) \, {\rm MHz}
\, ,
\end{equation}
so that the electrons which produce the radio emission we observe
have $\gamma \ga 10^4$ and short lifetimes $t_{IC} \la 2 \times 10^8$ yr.
Since these times are shorter than the durations of cluster mergers,
one expects to find radio emission only during the merger.
This result has been found previously based on more detailed models
(e.g., Sarazin 1999; 
Takizawa \& Naito 2000).

The total energy in cosmic ray electrons is then
\begin{equation} \label{eq:e_cr1}
E_{CR} \sim t_{IC} \dot{E}_{CR}
\propto \frac{f_{CR} f_{gas}}{( 1 + z )^{4}} \,
\left( \frac{M}{R} \right)^{5/2} \, . 
\end{equation}
It is also useful to replace the radius $R$, which is poorly defined
observationally, with the gas temperature $T$.
The gas in clusters is in hydrostatic equilibrium, which implies that
\begin{equation} \label{eq:tgas}
T \propto \sigma^2 \sim \frac{G M}{R} \, .
\end{equation}
The total energy of cosmic ray electrons then varies as
\begin{equation} \label{eq:e_cr2}
E_{CR}
\propto \frac{ f_{CR} f_{gas} }{ ( 1 + z )^4} \, T^{5/2} \, . 
\end{equation}
Alternatively, we can use the cluster X-ray luminosity $L_X$ to parameterize
the size of the cluster.
The X-ray luminosity--temperature relationship is approximately
$L_X \sim T^3$
(e.g., Arnaud \& Evrard 1999).
Thus, the energy in relativistic electrons varies as
\begin{equation} \label{eq:e_cr3}
E_{CR}
\propto \frac{ f_{CR} f_{gas} }{ ( 1 + z )^4} \, L_X^{5/6} \, . 
\end{equation}

The diffuse radio emission from a cluster depends both on the
population of relativistic electrons and on the magnetic field.
We consider two simple models for the variation in the field
strength from cluster to cluster.
First, we assume that the magnetic field is constant from cluster to
cluster.
In this case, electrons with the same energy or Lorentz factor produce
radio emission at the same frequencies in all clusters, and the total
magnetic energy density is the same in all clusters.
Then, the radio luminosity $L_{radio} \propto E_{cr}$, so that the
radio power at an emitted frequency $\nu$ varies as
\begin{equation} \label{eq:power_simple}
P_\nu \propto \dot{E}_{shock} \propto L_X^{5/6} \propto T^{5/2} \, .
\end{equation}
These relations are too flat to fit the observed relations
(Fig.~\ref{fig:P-L};
Feretti 2000; Liang et al.\ 2000).

As an alternative model, we assume that the magnetic field varies
from cluster to cluster in such a way that the magnetic pressure is
a fixed proportion of the gas pressure in the intracluster medium.
This might occur as a result of turbulence generated in merger
shocks.
Then, the magnetic field strength varies as
\begin{equation} \label{eq:B1}
B^2 \propto \rho_{gas} T \propto f_{gas} \, \frac{M}{R^3} \, T
\propto f_{gas} \, \frac{T^4}{M^2} \, .
\end{equation}
Assuming that clusters form from large scale structure and that the gas
is hydrostatic implies that the cluster mass and gas temperature are
related by $M \propto T^{3/2}$
(Bryan \& Norman 1998),
and this is consistent with the observations of clusters
(Horner, Mushotzky, \& Scharf 1999).
This implies that
\begin{equation} \label{eq:B2}
B \propto \left( f_{gas} T \right)^{1/2} \, .
\end{equation}

Varying the magnetic field also varies the energy of the electrons
which contribute to the radio emission at a given observing frequency
$\nu$ according to equation~(\ref{eq:nu_rad}), and this affects the
number of electrons effective at synchrotron emission at this frequency.
We assume that the electron population is a power-law in $\gamma$,
with $ N ( \gamma ) \, d \gamma = N_1 \gamma^{-p} \, d \gamma $
being the total number of electrons with energies in the range
$\gamma$ to $\gamma + d \gamma$.
We determine the power-law index $p$ from the observed
radio spectral index $\alpha$, which gives $p = 1 - 2 \alpha$.
Although the spectral indices vary from cluster to cluster, we assume
that there is no consistent variation with cluster X-ray luminosity
$L_X$ or temperature $T$.
We assume the electron populations is in steady-state, which implies that
the total number of particles or the normalization of the electron
spectrum varies as
\begin{equation} \label{eq:n1}
N_1 \propto \dot{E}_{CR} \propto f_{CR} f_{gas} T^{5/2} \, .
\end{equation}
 From synchrotron theory, the radio power at a frequency $\nu$ varies as
\begin{equation} \label{eq:power_sync}
P_\nu \propto \nu_B N \left( \nu / \nu_B \right)^{\alpha} \, ,
\end{equation}
where $\nu_B \equiv e B / ( 2 \pi m_e c) \propto B$ is the cyclotron frequency.
This leads to
\begin{equation} \label{eq:power_Bvary}
P_{\nu} \propto f_{CR} f_{gas}^{(3 - \alpha)/2} T^{3 - \alpha/2}
\propto L_X^{1 - \alpha / 6 } \, .
\end{equation}

These relations between radio power $P_{\nu}$ and $L_X$ or $T$ are
somewhat steeper than those produced by a constant magnetic field
(eq.~\ref{eq:power_simple}),
but are only a small improvement in terms of fitting the data.
We first tried varying $\alpha$, and found a best fit for $\alpha = -7$,
which is unreasonably steep.
We then fit the $L_X$--$P_\nu$ data in
Feretti (2000)
using this relation for several values of $\alpha$ that are consistent
with our data.
An example of these fits is shown in Figure~\ref{fig:P1.4-L}.
Since the slope is not very sensitive to $\alpha$, we show a fit for
only one typical value---other values give a similar fit.
This fit differs from the best fit by about 2.5$\sigma$.

\centerline{\null}
\vskip2.35truein
\includegraphics{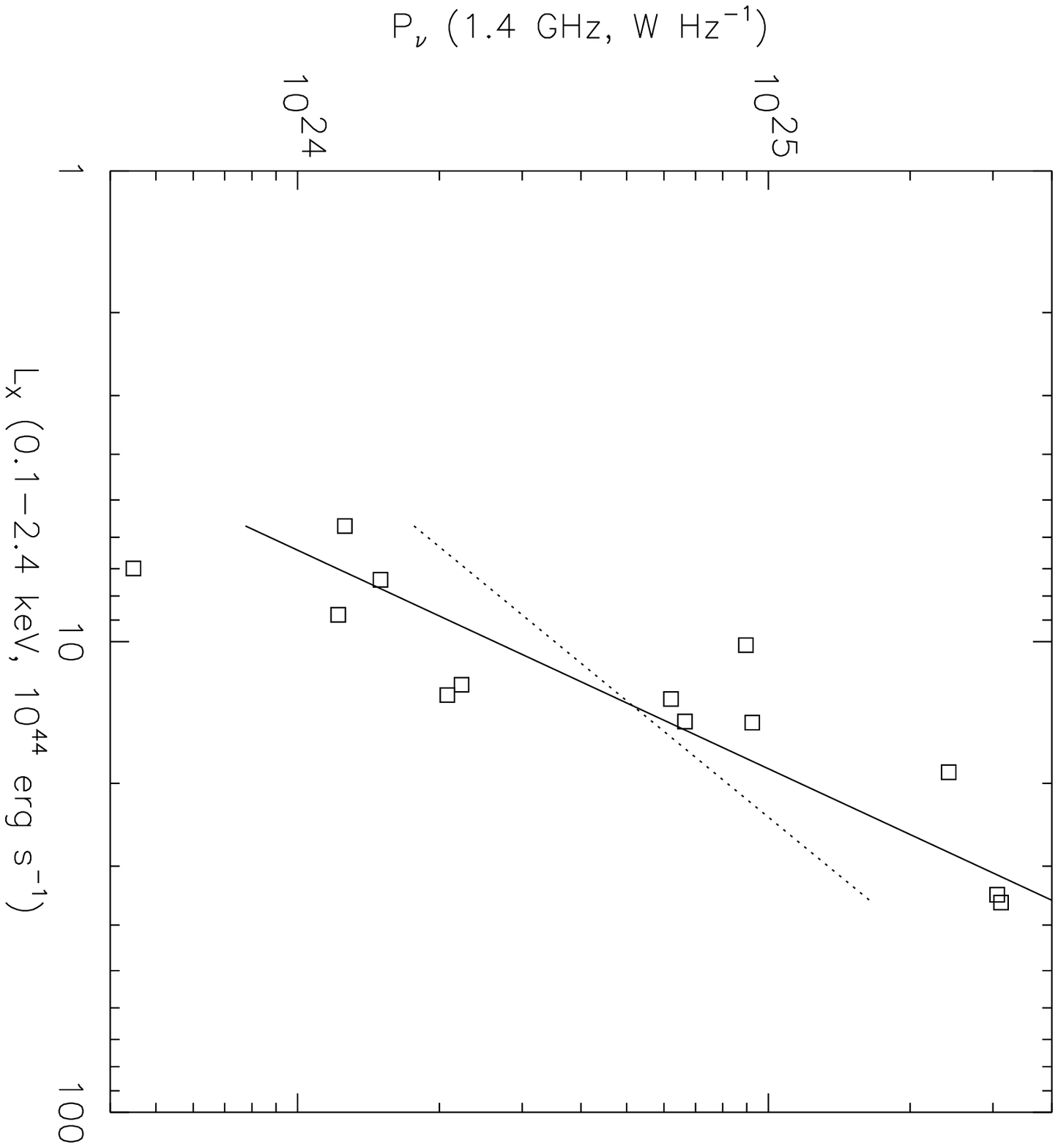}
\figcaption{Monochromatic power at 1.4~GHz of well-studied
radio halos from Feretti (2000) as a function of X-ray luminosity.
The solid line is a least squares fit to the data assuming a power-law
relation.
The dotted line is a fit using our simple shock model for $\alpha = -1.3$.
\label{fig:P1.4-L}}

For the purposes of considering the radio detectability of halos,
it is worth considering how the radio surface brightness $I_\nu$
and radio halo size might vary with X-ray luminosity or temperature.
The physical radius of a cluster is expected to vary with temperature
as $R \propto T^{1/2}$
(e.g., Mohr, Mathiesen, \& Evrard 1999).
If the temperature vs.\ X-ray luminosity relation is $L_X \propto T^3$
as assumed above, the cluster size varies as $R \propto L_X^{1/6}$.
If we assume that the size of the radio halo is proportional to the
cluster size $R$, then
$I_\nu \propto T^{3/2} \propto L_X^{1/2}$ for a fixed magnetic field,
and
$I_\nu \propto T^{ ( 4 - \alpha ) / 2} \propto L_X^{( 4 - \alpha ) / 6 }$
when the magnetic pressure increases in proportion to the gas pressure.
These relations are even flatter than those for the radio power.

\subsection{Limits from Clusters with Radio Non-Detections}
\label{sec:disc_nodetect}

The discussion so far has concentrated on the radio--X-ray correlations
for clusters with detected radio halos.
However, radio halos are relatively rare, and most clusters do not show
diffuse radio emission at a detectable level.
In a model where the
radio-emitting electrons are accelerated during
cluster mergers this is easily understood, since these electrons
have short lifetimes and will only be present during the period
of stronger merger hydrodynamical interaction.
Since clusters showing X-ray evidence for strong merger shocks are
also relatively rare and all radio halo clusters appear to be undergoing
mergers, this can explain, at least qualitatively, the low rate
of occurrence of radio halos and relics.

We also find that luminous halos and relics are generally not found in
clusters with low X-ray luminosities.
The NVSS radio survey at 20~cm found a similar result
(Giovannini et al.\ 1999).
This relationship between X-ray luminosity and the detection of a halo or
relic appears in several other forms, as well.
The Abell clusters north of $\delta=30^{\circ}$ that fall into the
Ebeling et al.\ (1996) X-ray--bright cluster sample comprise less than
20\% of our initial sample, but make up 65\% of the clusters in which we
find diffuse emission.
If X-ray luminosity and the radio halo luminosity were uncorrelated and
the halo detection rate from the Ebeling et al.\ (1996) sample held for
fainter clusters, we would expect to find a radio halo or relic in about
200 of the clusters we studied.

Surely, some of the low X-ray luminosity clusters in the Abell catalog
are also undergoing mergers, and may have particle acceleration.
Why are these lower luminosity clusters not detected in radio?
The observations of clusters indicate that there is a steep correlation
of radio power with X-ray luminosity or temperature;
our simple merger shock model also implies a steep correlation, although
probably not as steep as the data.
Here, we consider the possibility that the failure to detect radio
halos in low X-ray luminosity clusters results from this correlation
and the radio sensitivity of our survey.
We assume that the correlation between radio halo and X-ray luminosity
seen for bright clusters
(Feretti 2000) continues to lower X-ray luminosities.

The surface brightness sensitivity limit of our search technique is
taken to be that of the WENSS, shown in
Figure~\ref{fig:sensitivity_sb}.
We determine the predicted surface brightnesses of clusters as
a function of X-ray luminosity $L_X$ and redshift.
For a cluster with a given X-ray luminosity, we used the
radio power vs.\ X-ray luminosity relation
(Figure~\ref{fig:P-L})
to determine its radio power at 327~MHz.

To determine the predicted mean radio surface brightness of the halo,
we also need to know its physical size.
It seems likely that that bigger clusters, which have
larger X-ray luminosities, will also have larger radio halos.
For example, it may be that radio halo sizes are proportional to
the overall sizes of clusters.
As discussed in \S~\ref{sec:disc_model}, this leads to the
radio halo size varying with X-ray luminosity roughly as
$R \propto L_X^{1/6}$.
In our sample of radio halos, there is no clear evidence for
a variation of the size of the halo with the X-ray luminosity of
the cluster.
However, it may be that our sample is too small, or that the
sizes deduced from the WENSS images are too uncertain.
Figure~\ref{fig:Size-L} plots the sizes of the radio halos from
the sample in
Feretti (2000)
as a function of the cluster X-ray luminosity.
This sample does show some evidence for a size-luminosity relationship.
The dashed curve is the best-fit relation which follows the
cluster radius vs.\ X-ray luminosity relation, $R \propto L_X^{1/6}$.
The solid curve is the best-fit power-law relation with an
arbitrary exponent.
This best-fit corresponds roughly to $R \propto L_X^{1/2}$.

\centerline{\null}
\vskip2.55truein
\includegraphics{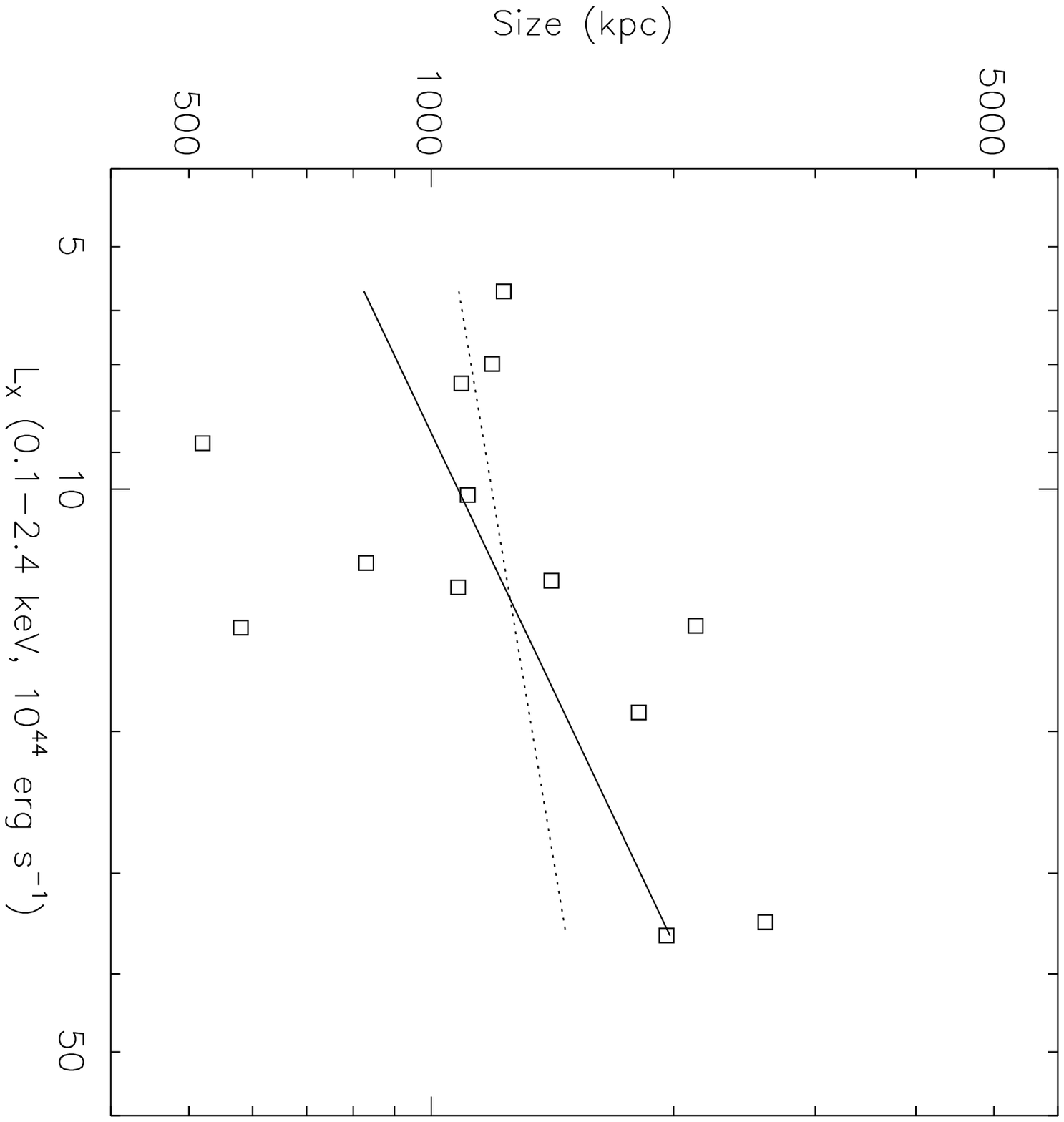}
\figcaption{Largest linear radio size vs.\ X-ray luminosity for
the halos in Feretti (2000).
The solid line is the best least squares fit to the data for a power-law
relationship.
The dotted line is the best fit with an exponent of $1/6$, which would
apply if the radio halo size was proportional to the cluster size.
\label{fig:Size-L}}
\vskip0.2truein

The sizes shown in Figure~\ref{fig:Size-L} are largest linear size
($LLS$) values.
Typically, the regions covered by the radio halos are not circular.
For the halo sources in our survey, the average area of the source
(the area used to determine its mean surface brightness in
Figure~\ref{fig:sensitivity_sb}) is given by
$0.37 ( LLS )^2$.
We use the same factor to determine the predicted area of a source
of a given X-ray luminosity;
its mean surface brightness at 327 MHz in the cluster rest frame is
then determined by dividing the predicted radio power $P_{327}$ by
the predicted area of the halo.
For the observed radio power and radio size vs.\ X-ray luminosity relations,
the predicted mean radio surface brightness is a monotonically increasing
function of the X-ray luminosity.

For a cluster at a redshift of $z$, the observed surface brightness
at an observing frequency of 327 MHz is reduced by a factor of
$( 1 + z )^{-3 + \alpha}$.
We assume a typical spectral index of $\alpha = -1.3$ to
determine the observed surface brightness.
This observed mean surface brightness was compared to the surface
brightness limit of our survey (Figure~\ref{fig:sensitivity_sb}), and
the X-ray luminosity of the faintest detectable cluster was determined.
This limiting $L_X$ is plotted versus redshift as the solid curve in
Figure~\ref{fig:L-z}.
This limit is incorrect for very small redshifts
where the halos may be too large in angular size, and at very large
redshift where the halos might be unresolved.
The sense of these differences is to make the clusters harder to detect.

It is clear that the detections are roughly consistent with the survey
sensitivity and the radio--X-ray correlations.
Abell~2255 is the only cluster to lie at or below the limit,
but as Figure~\ref{fig:P-L} shows, this cluster deviates
substantially from the best-fit $L_X$--$P_{\nu}$ relation with a low
X-ray luminosity given its radio power.
It is clear that radio halos in fainter X-ray clusters
($L_X \la 5 \times 10^{44}$ ergs s$^{-1}$)
would be too weak to be observed, if they follow the radio--X-ray
correlations observed for brighter clusters.
Of course, there may be other selection effects affecting the detection
rates of clusters.
For example, some of the radio halo detections have resulted from
deep radio observations of Sunyaev-Zel'dovich clusters to remove
radio sources
(e.g., Moffet \& Birkinshaw 1989; Liang et al.\ 2000).
Since the S-Z clusters tend to be selected as the hottest, highest
X-ray luminosity clusters, this might explain part of the correlation
of radio halo detections with cluster X-ray temperature or luminosity.

\centerline{\null}
\vskip2.55truein
\includegraphics{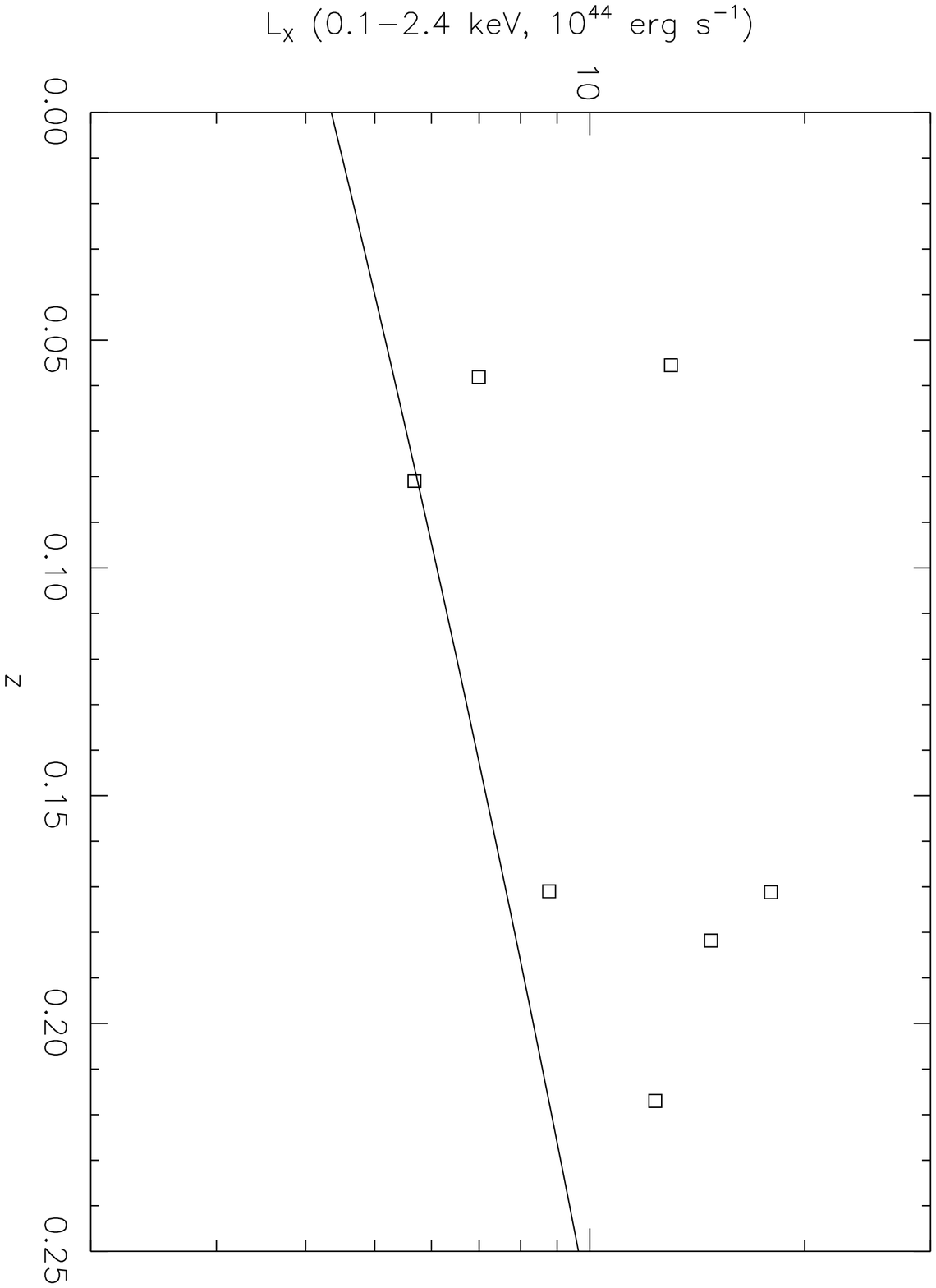}
\figcaption[fig7.eps]{X-ray luminosity of halo clusters in the 0.1--2.4
keV rest frame as a function of redshift.
The solid curve gives the lower limit on cluster X-ray luminosity for
a detection of a halo with $\alpha = -1.3$ in the WENSS survey,
assuming the halo radio power and size vary with X-ray luminosity
as shown in Figures~\protect\ref{fig:P-L} and \protect\ref{fig:Size-L}.
\label{fig:L-z}}

\section{Conclusions} \label{sec:concl}

We have discovered 7 new candidate radio halos and relics in a search of
Abell clusters in the WENSS.
We also confirm the presence of diffuse emission in 7 clusters and find
further evidence of such emission in 2 more clusters where the presence
of a halo has been posited but remains uncertain.
Our search technique detected all but one previously known source that
fell within our sample.
More detailed radio observations of the new radio halos and relics would
be very useful to accurately determine their structure.

We argue that radio halos or relics are only found in clusters which
are currently undergoing a cluster merger.
All the clusters in our sample show evidence for a merger from either
their X-ray surface brightness distribution or their galaxy
distribution.
This can explain the relative rarity of diffuse radio emission in
clusters.

We also find weak evidence for the observed correlation between
monochromatic power of radio halos and cluster X-ray luminosity.
We present a very simple model for the correlation of radio power
with X-ray luminosity or temperature in clusters which are currently
merging, on the assumption that the radio-emitting electrons are
accelerated by merger shocks.
We consider two cases for this model: one in which the magnetic field is
the same for all clusters and one in which the field varies as a fixed
proportion of the gas pressure.
The latter model is marginally more successful at fitting the data.
This argument predicts a strong radio--X-ray luminosity
correlation, although not as steep as the one observed at 20~cm.

Our survey is the first to look for halos in a large sample of clusters
with low X-ray luminosity.
In general, radio halos and relics are not found in low X-ray luminosity
clusters.
We argue that this is the result of the steep radio power vs.\ X-ray
luminosity correlation.
If this is true, many more low luminosity clusters could be detected
as diffuse radio sources if the sensitivities of the surveys could be
greatly increased.
Clearly, deeper imaging with high sensitivity at short ($u,v$) spacings
is necessary to  test whether the $P-L_X$ correlation holds  at lower
X-ray luminosities.

\acknowledgements
We thank Bill Cotton for many helpful discussions regarding the WENSS.
We are also grateful to the referee for several helpful suggestions.
Support for this work was provided by the National Aeronautics and Space
Administration through {\it Chandra} Award Numbers GO0-1019X, GO0-1141X,
and GO0-1173X
issued by the {\it Chandra} X-ray Observatory Center,
which is operated by the Smithsonian
Astrophysical Observatory for and on behalf of NASA under contract
NAS8-39073.

\setcounter{figure}{1}
\begin{figure*}[p]
\begin{center}
\epsfig{file=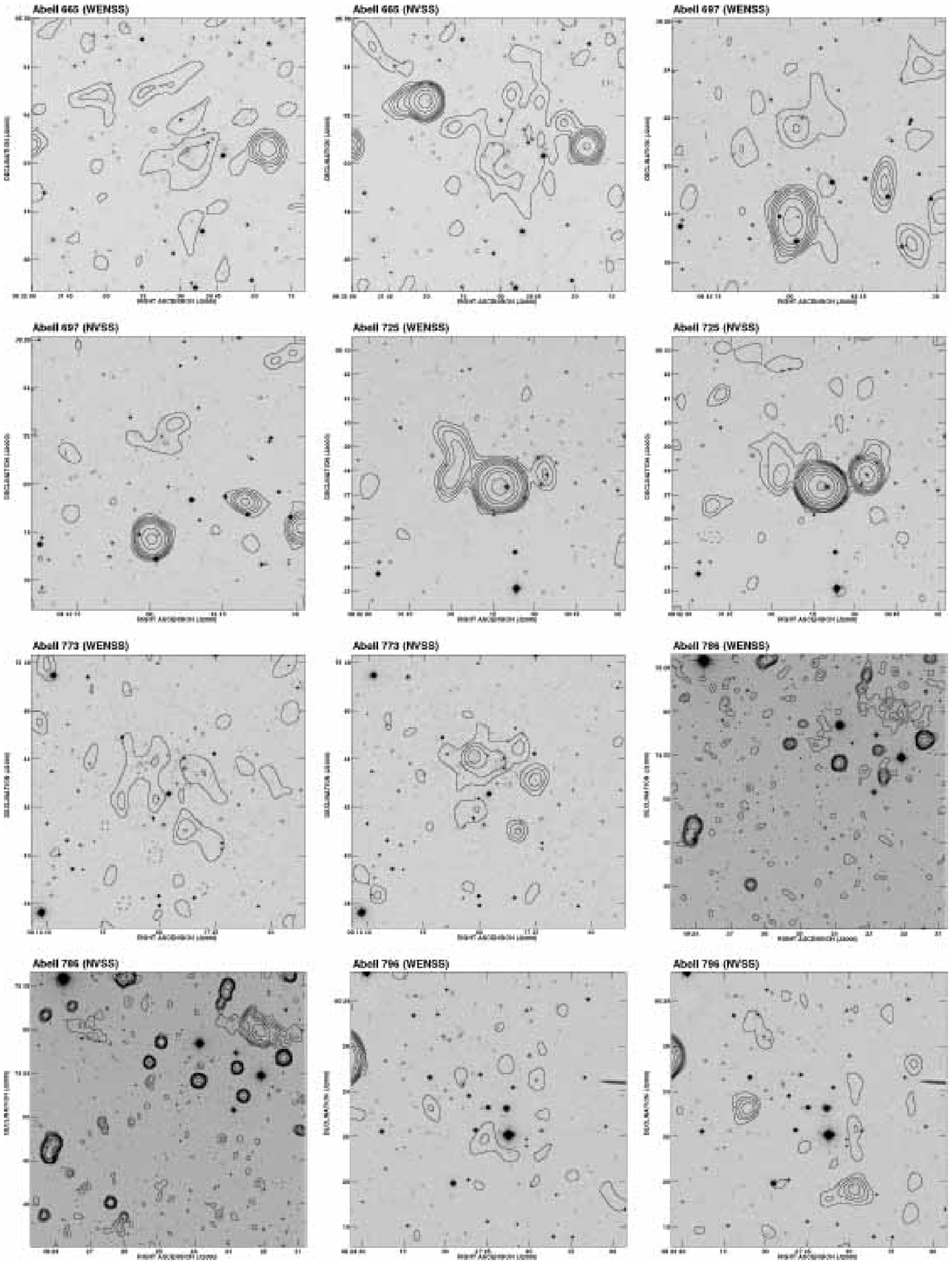, height=8.8in}
\end{center}
\caption{\footnotesize For each cluster in
Table~\protect\ref{tab:candidates}, contours from the WENSS and NVSS
radio images are given, superposed on the DSS optical images.  The radio
contours shown are 2, 3.5, 5, 7, 10, 15, 30, 50, 100, 200, 500, \& 1000
times the local rms. The 3.5$\sigma$ negative contour is also shown.
Typical rms levels are 3.6 mJy beam$^{-1}$.  Most of the images are
approximately 12\arcmin$\times$12\arcmin; all include the cluster
center.
\label{fig:images}}
\end{figure*}

\begin{figure*}[p]
\begin{center}
\epsfig{file=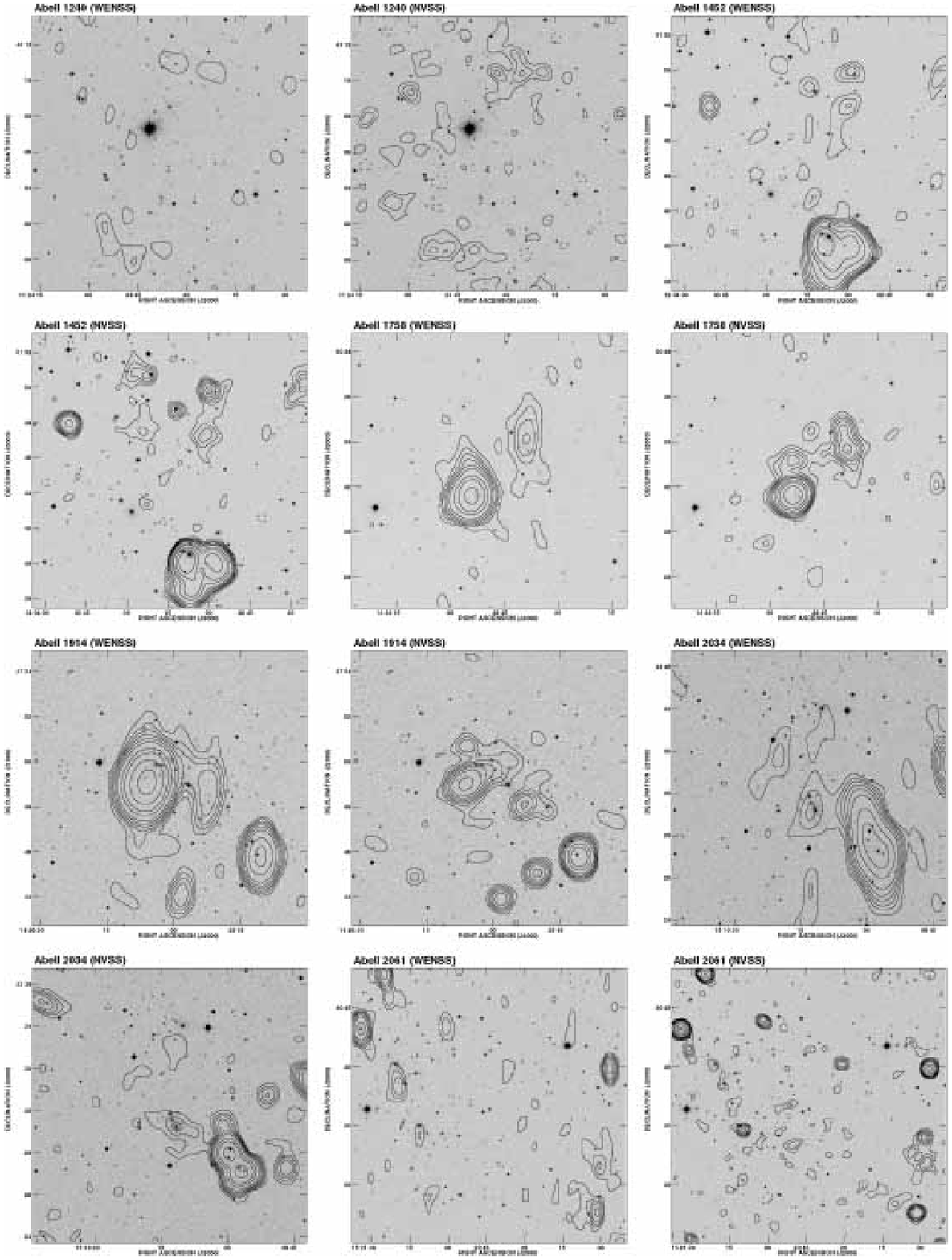, width=6.5in}
\end{center}

{\footnotesize \sc {Fig. 2} cont'd}
\end{figure*}

\begin{figure*}[p]
\begin{center}
\epsfig{file=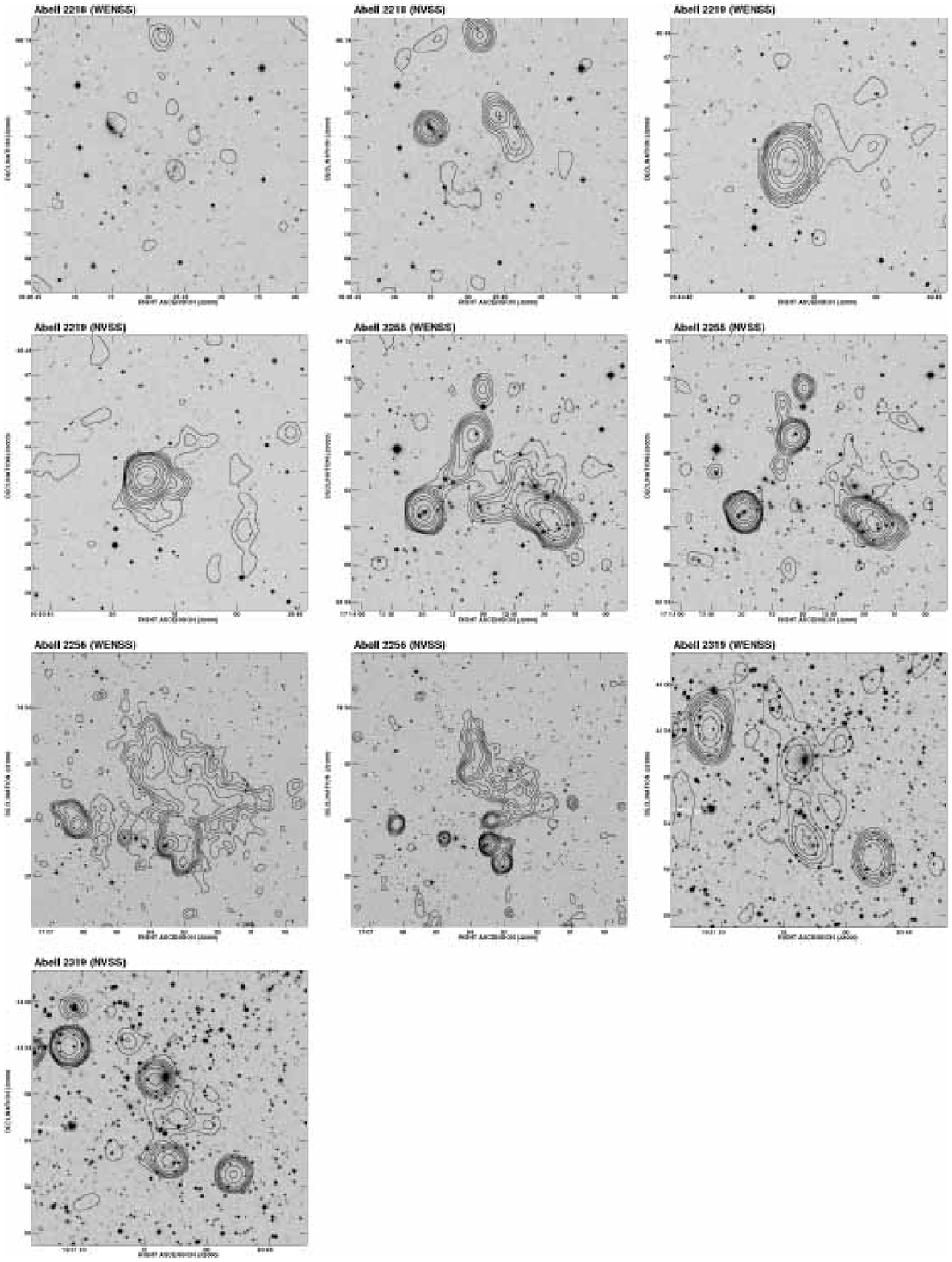, width=6.5in}
\end{center}

{\footnotesize {\sc Fig. 2} cont'd}
\end{figure*}

\end{document}